**Single photon emission from ZnO nanoparticles**

Sumin Choi[1], Brett C. Johnson[2] Stefania Castelletto[3], Cuong Ton-That[1], Matthew R. Phillips[1] and Igor Aharonovich[1]*

[1] School of Physics and Advanced Materials, University of Technology Sydney, Ultimo, New South Wales 2007, Australia
[2] Centre for Quantum Computation and Communication Technology, School of Physics, University of Melbourne, Victoria 3010, Australia
[3] School of Aerospace, Mechanical and Manufacturing Engineering RMIT University, Melbourne, Victoria 3000, Australia

* Email: igor.aharonovich@uts.edu.au

**Abstract**

Room temperature single photon emitters are very important resources for photonics and emerging quantum technologies. In this work we study single photon emission from defect centers in 20 nm zinc oxide (ZnO) nanoparticles. The emitters exhibit bright broadband fluorescence in the red spectral range centered at 640 nm with polarized excitation and emission. The studied emitters showed continuous blinking, however, bleaching can be suppressed using a polymethyl methacrylate (PMMA) coating. Furthermore, hydrogen termination increased the density of single photon emitters. Our results will contribute to the identification of quantum systems in ZnO.

Sources of non-classical light, are important for a range of applications in quantum communications, sensing and information processing[1, 2]. A particular emphasis is placed on the development of solid-state single photon sources that operate at room temperature, as those are the most promising sources for scalable and integrated devices. Color centers in diamond[3], and in particular the nitrogen vacancy[4] and the silicon vacancy defects[5, 6], have been the subject of intense study in recent years due to their unprecedented photostability and spin properties. Recently, several additional materials have emerged as candidates for single photon generation, including site controlled GaN quantum dots[7], defects in silicon carbide[8] and zinc oxide (ZnO)[9]. While the single photon emission in SiC is attributed to the intrinsic defect known as an "antisite – vacancy pair", the origin of the quantum emission in ZnO remains under debate.

ZnO offers an interesting and valuable system to study quantum effects at the nanoscale[10]. ZnO nanoparticles are commercially available while nanowires and other nanostructures can be easily synthesized by hydrothermal[11] or chemical vapour deposition growth techniques[12, 13]. In addition, ZnO is very suitable for photonic applications, as it has a wide, direct bandgap (3.4 eV) and a large number of intrinsic defects and impurities that emit from the ultra-violet to the near infrared[14, 15]. Finally, photonic elements including microdisk cavities[11, 16, 17] can be easily fabricated from this material. Therefore, there is an urgency to understand and fully characterize the photophysical properties of single emitters to transform ZnO as a valuable platform for quantum photonic applications.

In this letter we study single photon emission from point defects in ZnO nanoparticles emitting at ~ 640 nm. We perform polarization measurements and study the effect of surface termination on the emission properties of these single emitters.

ZnO nanoparticles (average size 20 nm, Nanostructured and Amorphous Materials Inc.) were spin coated onto a silicon substrate (0.5 cm x 0.5 cm). Then, the substrates were annealed at 500°C in air using a horizontal tube furnace to remove graphitic residue in the powder and to induce the formation of color centers that produced a bright red fluorescence. Single photon emission characteristics were investigated at room temperature under ambient conditions using scanning confocal microscopy. Briefly, a continuous wave pump laser with a wavelength of 532 nm was used as an excitation source, and was focused on the sample through a high (0.9) numerical aperture objective lens (Nikon, x100). The signal was collected using the same objective, directed through a dichroic mirror and a long pass filter (Semrock) and coupled into a multimode fibre (62.5 μm core) that acts as a confocal aperture. The confocal excitation spot in our setup is estimated to be ~ 600 nm. Single photon emission was measured using a standard Hanbury-Brown & Twiss (HBT) setup, while the photoluminescence (PL) spectra were collected using a Princeton Instruments spectrometer with 300 mm focal length.

Fig. 1(a) shows a representative secondary electron scanning electron microscopy (SEM) image of the ZnO nanopowder after annealing in air, where the nanoparticles are clearly visible. Fig. 1(b) shows a typical confocal map of the ZnO nanopowder using a 532 nm excitation laser. Several bright spots (marked as circles in Fig. 1(b)) are visible, and correspond to localized bright fluorescent defects.

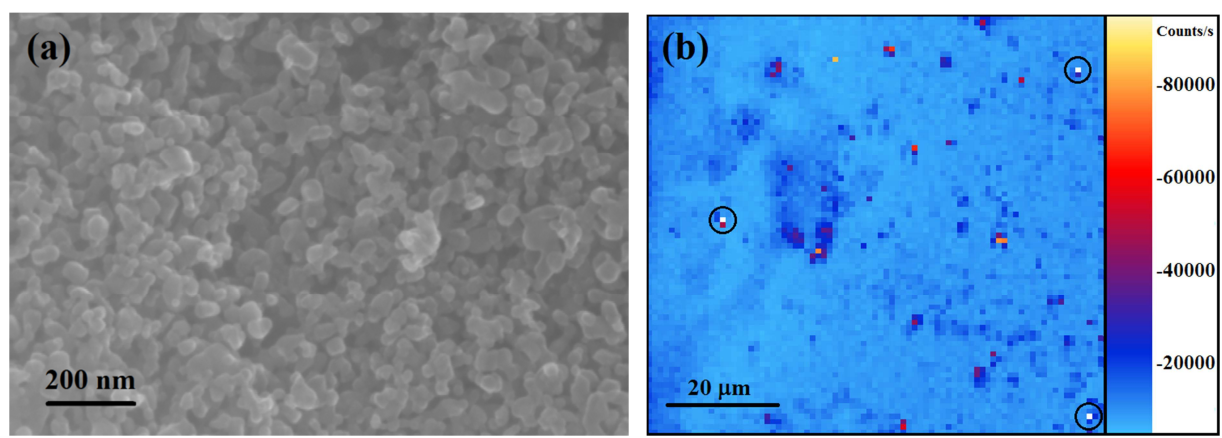

**Figure 1. (a) SEM image of a dense film of ZnO nanoparticles annealed at 500 °C in air. (b) An 80 x 80 μm² scanning confocal image of ZnO nanoparticles excited with a 532 nm laser. Bright isolated emitters are indicated with the circles.**

Fig. 2(a) shows a room temperature PL spectrum recorded from one of the bright spots in the confocal map shown in Fig. 1b (black curve). The curve shows a red luminescence peak at around 640 nm which can be potentially attributed to zinc vacancies ($V_{zn}$)[18]. The background fluorescence exhibited a broad weak red luminescence (red curve, Fig. 1b). Fig. 2(b) shows the second order correlation function, $g^2(\tau)=<I(t)I(t+\tau)>/<I(t)>^2$, that describes the probability of detecting a photon at a delay time, $\tau$, given that a photon was detected at time $t$, recorded from the same emitter using the HBT configuration. The pronounced antibunching dip in the photon statistics at zero delay time ($g^2(0) \sim 0.1$) indicates that the emission originates from a single photon emitter. The deviation from 0 is attributed to the overall background luminescence. The photon statistics of the single emitters can be described with a frame of a two or three level system[9, 19]. For the three level systems, at higher excitation powers, bunching behavior is observed, indicating the presence of a metastable (shelving) state. Only 10 % of the investigated emitters showed strong bunching behavior. Emission from those defects was also associated with a very narrow emission line (~ 5 nm at full width at half maximum (see supplementary information[20]). Most of the emitters, however, exhibited moderate or no evidence of a shelving state.

The lifetime of the emitters can be deduced from fitting the second order correlation function to the following equation, $g^2(\tau)=1-exp(-\lambda_1/\tau)$, at low excitation powers, where $\lambda_1$ is the decay rate of the emitter. The lifetime of the investigated emitter is therefore 6.4 ns – comparable with other single defects in solids.[3, 9]

The saturation curve of a single emitter determined by summing the count rates output by the two APDs of the HBT setup is shown in Fig. 2(c). The emission count rate as a function of excitation power is fit using the following equation:

$$\emptyset = \frac{\emptyset_\infty P_{opt}}{P_{sat}+P_{opt}} \quad (1)$$

where $\emptyset$ represents the single-photon count rate at a given excitation power ($P_{opt}$), $\emptyset_\infty$ is the saturation count rate at optical saturation power ($P_{sat}$). From this fit, the saturation count rate was determined to be $1.84 \times 10^5$ counts/s with an optical saturation power of 1.46 mW. Some emitters were brighter with count rates approaching ~ $10^6$ counts/s. The photon statistics of the ZnO defect and its brightness are comparable with other solid state room temperature single photon emitters.

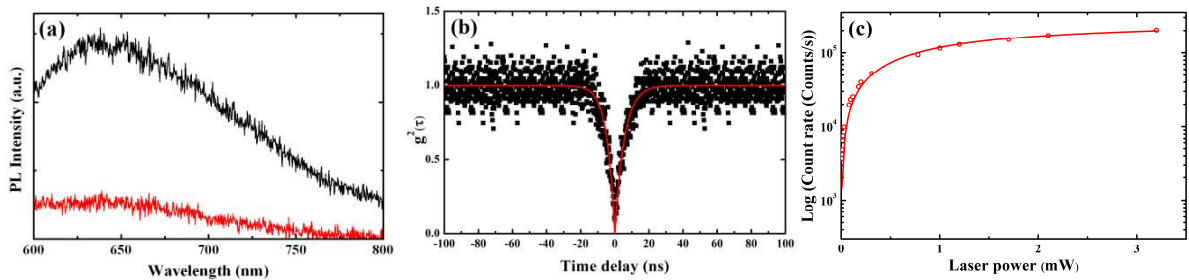

**Figure 2. (a) PL spectrum recorded from a bright defect within the ZnO nanoparticles (black curve). The red curve represents the background emission. (b) The presence of a single quantum emitter in the ZnO nanoparticles is revealed by the second-order autocorrelation function, $g^{(2)}(\tau)$, corresponding to the PL in Fig 2(a). The dip at zero delay time indicates that it is a single emitter. The solid line is a fit using a two level model. (c) Single photon emission count rate versus excitation power. The red circles are the raw data and the solid line is the fit using Eq. 1. The saturation count rate for this emitter is $1.84 \times 10^5$ counts/s.**

The excitation polarization behaviour of the single photon emission was measured by rotating a half-wave plate in the excitation path, and the result is shown in Fig. 3 (red curve). The emission polarization was characterized by fixing the polarization excitation to a maximum and rotating the polarizer at the collection channel. An almost full extinction of the emission was observed (Fig. 3 blue curve), indicating that emission occurs from a linearly polarized dipole. The mismatch between the polarization and the emission dipoles originates from the redistribution of the electrons upon photon absorption.

The polarization visibility (defined as $V= (I_{max}-I_{min})/(I_{max}+I_{min})$, where $I_{max}$ and $I_{min}$ are maximum and minimum intensities, respectively) of the emitter is 85 %. High visibility is very important for employing sources for quantum communications as the information is encoded in the polarization state of the emitter. Therefore, highly polarized emitters can provide excellent signal to noise ratios.

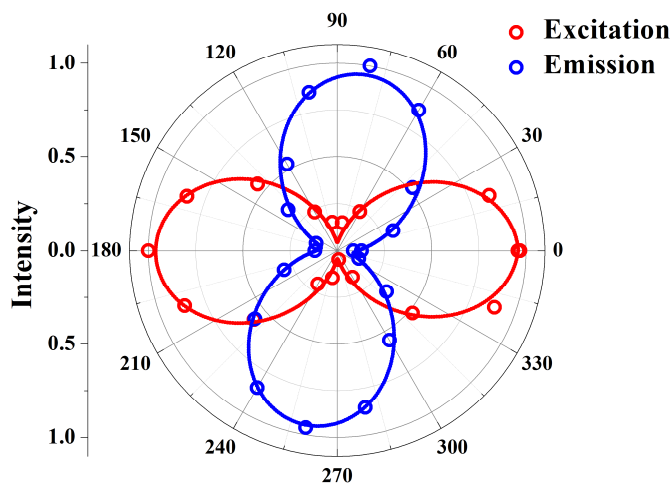

**Figure 3. Polarization measurement of a ZnO single emitter for both polarized excitation (red curve) and emission (blue curve) as a function of the polarizer angle. The circles are the raw data and the solid line is a fit following Malus' law.**

All of the investigated emitters in our experiments showed blinking and eventual bleaching after several minutes. To investigate the bleaching behaviour, we performed two treatments to modify the ZnO surface. In the first experiment, the ZnO nanoparticles were exposed to an hydrogen plasma (2 minutes, 15 W, 473K). The hydrogen plasma treatment leads to quenching of the red luminescence due to the passivation of acceptor–like defects, possibly the $V_{zn}$.[21] The reduction in red luminescence is confirmed by cathodoluminescence measurements on the ZnO nanoparticles (see supplementary information[20]). The second process involved the deposition of 200 nm of optically transparent polymethyl methacrylate (PMMA) to coat the nanoparticles. This was inspired by recent reports on the quenching suppression of emission in diamond nanoparticles[22] using polymer coatings. After each step, confocal maps were recorded and single emitters were identified. Fig. 4a,b show antibunching curves recorded from single emitters after hydrogen plasma and after PMMA coating, respectively. It is clear that the emitters still maintain their quantum behaviour after the hydrogen passivation treatment and PMMA coating.

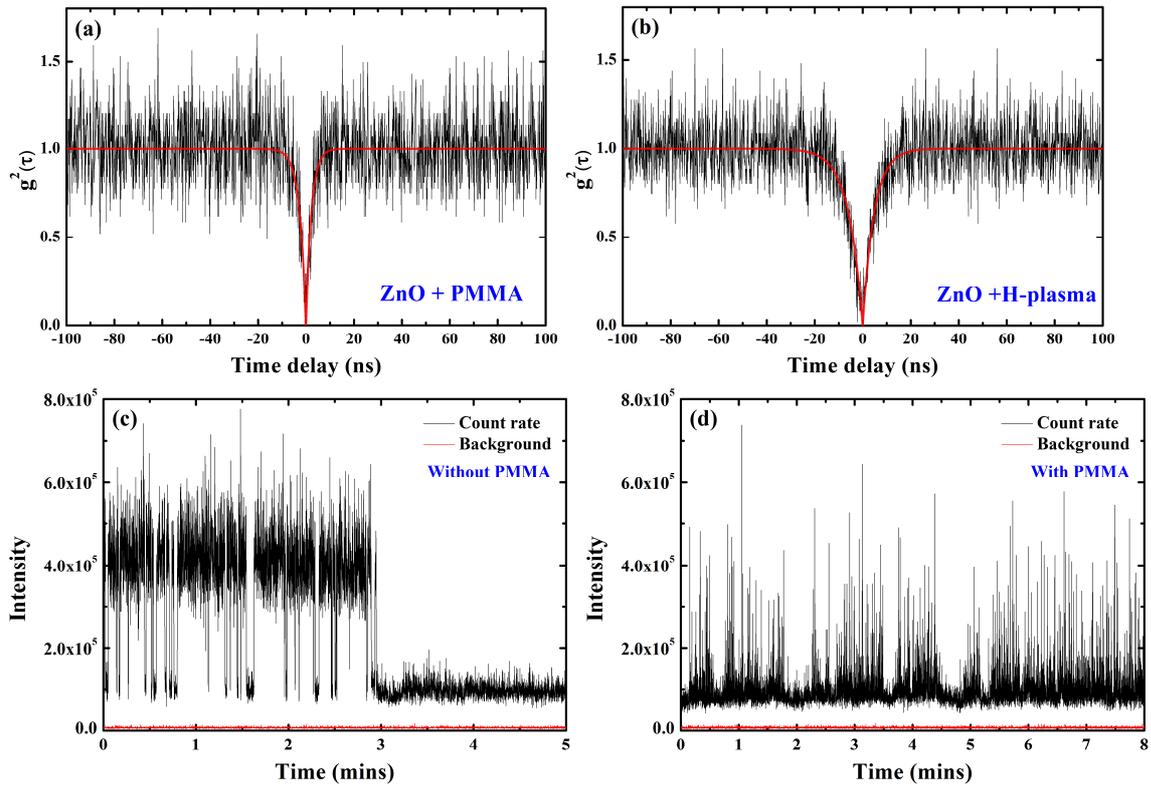

**Figure 4.** Second order correlation functions recorded from single defects within ZnO after (a) coating with PMMA and (b) exposure to hydrogen plasma. (c) Intensity trace recorded from a single emitter within an untreated ZnO nanoparticle showing permanent bleaching after 3 minutes. (d) Intensity trace of a single emitter of a ZnO nanoparticle coated with PMMA that exhibits blinking but no bleaching. The red curve in (c,d) is the background florescence.

Fig. 4 (c,d) show the defect stability recorded from a blinking single emitter from untreated nanoparticles and PMMA coated nanoparticles, respectively. The blinking traces clearly show that the emitters embedded in ZnO nanoparticles that have only been annealed, bleach after several minutes of excitation with a laser beam. Over 20 emitters were measured and all exhibited equivalent behaviour. Conversely, the ZnO nanoparticles that are coated with PMMA exhibited blinking, but did not bleach under same excitation conditions. The time constants associated with the on/off PL intermittency follows an exponential trend, in contrast to that observed for quantum dots, where a power law is followed, but similar to the behaviour of other point defects in semiconductors.[8, 22, 23] The mechanism giving rise to this blinking may likewise arise from a photo-induced charge conversion of the defect, to a charge exchange with other impurities in close proximity that act as charge traps or from low probability transitions to a meta-stable dark state. The duration at which the system spends in the "off" state is in the range 70 – 160 ms.

Hydrogen passivation did not yield similar results, and the investigated emitters exhibited similar bleaching as the untreated ZnO nanoparticles. However, after exposure to the hydrogen plasma, the density of emitters increased by a factor of three compared with the untreated samples. The PMMA coating did not influence the absolute number of observed single defects. This indicates the likelihood that the quantum emitters are charged defects, which are influenced by the surface termination and can be switched on and off[24, 25]. Hydrogen termination is likely to induce band bending, thus stabilizing or increasing the probability to populate the single emitter. Hydrogen is also rapidly diffuses in ZnO nanoparticles[26], and can create local charge traps and defects that increase the formation of quantum emitters. Furthermore, during the "off" and bleaching states, the emitters are still bright (~ 80,000 counts/s) compared to the background fluorescence from the particles (~ 6000 counts/s – marked as a red base line in Fig. 3(c,d). However, this emission does not exhibit quantum behaviour. This may suggest that the quantum emitter is likely to be a different charge state of the same broad emission that contributes to bright fluorescence or associated to transition metal impurities that do have emission at a similar spectral range[27]. At this point, however, we are not able to assign the chemical origin of the quantum emission to a specific defect center.

In summary, single photon emission from point defects in 20 nm ZnO nanoparticles was observed. The quantum emitters can easily be formed through annealing in air – providing an excellent platform for easy access to solid state quantum systems at room temperature. The emitters exhibit red fluorescence centered at 640 nm with a high count rate (~200,000 counts/s) and display a fully polarized emission. The as-grown and annealed emitters all exhibit blinking which quenched after exposure to the laser beam after several minutes. However, no bleaching was observed after the ZnO nanoparticles were coated with PMMA. Exposure to a hydrogen plasma treatment increased the yield of the single photon emitters. Further studies to optimise the formation probability of these quantum emitters are required. Finally, generation of these quantum emitters in bulk material is important to broaden their applicability to various quantum applications.

**Acknowledgements**

I.A. is the recipient of an Australian Research Council Discovery Early Career Research Award (Project No.DE130100592). B.C.J. acknowledges the Australian Research Council Centre for Quantum Computation and Communication Technology (CE110001027). Support from the Australian Research Council Grant DP0986951 is gratefully acknowledged. The authors would like to thank G. McCredie for technical support.